\documentclass[aps,preprint]{revtex4}%
\usepackage{amsfonts}
\usepackage{amsmath}
\usepackage{amssymb}
\usepackage{graphicx}%
\providecommand{\U}[1]{\protect\rule{.1in}{.1in}}

\begin{document}
\title[ ]{Conformal Symmetry and the Thermal Effects of Acceleration in Classical Physics}
\author{Timothy H. Boyer}
\affiliation{Department of Physics, City College of the City University of New York, New
York, NY, USA 10031-9198}
\email{tboyer@ccny.cuny.edu}
\keywords{}
\pacs{}

\begin{abstract}
An accelerating Rindler frame in Minkowski spacetime acting for a finite time
interval is used to carry a box of particles or waves between two relativistic
inertial frames. \ The finite spatial extent of the box allows treatment of
the equations of motion for particles or for waves, while the Rindler
acceleration provides a substitute for scattering to test for thermal
equilibrium. \ In the case of equilibrium for relativistic particles, the
J\"{u}ttner distribution is derived. \ For relativistic waves, a full
derivation of the Planck spectrum including zero-point radiation is obtained
within classical theory. \ For relativistic waves, relativistic behavior and
conformal symmetry are crucial. \ It is emphasized that the classical
two-point correlation function for classical zero-point radiation depends upon
the geodesic separation between the spacetime points and is independent of the
coordinate system choice. \ The classical point of view here does not give any
support for the idea that a system in uniform acceleration through classical
zero-point radiation finds a thermal spectrum. \ 

\end{abstract}
\maketitle

\section{Introduction}

In classical theories with random distributions of particles and waves, we are
often interested in \textquotedblleft equilibrium\textquotedblright%
\ situations. \ If we consider a box with perfectly reflecting walls for the
particles and waves in an inertial frame, the distribution of particle speeds
or the spectrum of normal wave modes will not change with time. \ The idea of
equilibrium requires that the averages over the \textquotedblleft
equilibrium\textquotedblright\ distribution will not be changed by the
introduction of a scattering system which changes the particle velocities or
spectrum of wave normal modes. \ For example, a thermal distribution of waves
at temperature $T$ is supposed to be unchanged under scattering by a
\textquotedblleft black particle.\textquotedblright

\ A theoretical account of the scattering \textit{will depend upon the theory}
which is used in the calculation of the scattering. \ Mixtures of theories
will give results dependent upon the reference frame chosen. \ For
two-particle scattering, for example, if one uses\textit{ nonrelativistic
}energy and momentum expressions for one particle but uses\textit{
relativistic energy and momentum }for the second\textit{ }particle, then
the\textit{ }result of a collision will depend upon the\textit{ inertial frame
in which the collision is analyzed.}\cite{B2009}\textit{ \ }Also, for
a\textit{ charged }particle undergoing \textit{acceleration} (and so
radiating)\textit{ }and also interacting with electromagnetic waves which
\textit{balance }the particle's loss of energy to electromagnetic
radiation,\textit{ only the Coulomb potential }for the particle motion will
give the same result in all \textit{relativistic} inertial
frames.\cite{B2023d}\textit{ }\ Thus one must be careful when analyzing
\textquotedblleft equilibrium\textquotedblright\ to specify what theory is
being used in the analysis of the scattering, and one must use that same
theory for \textit{all} aspects of the analysis. \ 

In the present article, we emphasize \textit{relativity} and \textit{conformal
symmetry} when analyzing the stability of distributions of either free
\textit{relativistic} particles or source-free \textit{relativistic} waves in
a box with perfectly reflecting walls. \ Of course, in an inertial frame,
neither free particles nor source-free waves will experience a meaningful
scattering which determines equilibrium. The \textit{scattering} aspect in
this article is provided by moving the box from one inertial frame to any
other inertial frame with a nonvanishing relative velocity $\mathbf{V}$
between the frames. \ In order to carry out this transfer between inertial
frames with relative velocities $\mathbf{V}$, the box must be accelerated.
\ When one compares the distribution in the initial inertial frame before the
acceleration began and the distribution in the final inertial frame after the
acceleration ceases, equilibrium requires that the average values
corresponding to equilibrium have not changed for any final inertial frame.
\ Thus \textit{acceleration} in a box with perfectly reflecting walls provides
the \textquotedblleft scattering\textquotedblright\ aspect for the analysis of
equilibrium. \ We require that the random distribution is stable in
\textit{both} an inertia and also in a Rindler frame. \ This stability is our
test for equilibrium. \ 

The use of acceleration in connection with equilibrium for \textit{non}%
relativistic particles goes back to the analysis of Maxwell and
Boltzmann.\cite{Boltzmann} \ Here we extend this acceleration analysis to
\textit{relativistic particles} and \textit{relativistic waves }within
classical physics\textit{. \ }A \textit{Rindler frame} acting for a
\textit{finite time} is precisely the needed aspect for the accelerations of
relativistic theory in flat spacetime. \ It is this \textit{relativistic
}aspect\textit{ }which is often missed in the analysis of thermal radiation.
\ In current physics, this aspect is missing from the textbooks of modern
physics and in all the treatments which use a \textit{nonrelativistic} charged
harmonic oscillator to scatter \textit{relativistic} waves. \ 

The other aspect which is often unappreciated is that of conformal symmetry.
\ If a charged particle is in circular motion in a potential, the charge will
radiation and so lose energy. \ The energy situation may be stabilized by
introducing two counter-propagating plane electromagnetic waves along the
symmetry axis of the circular orbit. \ However, the only potential which will
meet the requirements of Lorentz invariance is the Coulomb potential for the
circular orbit.[2] \ The relativistic Coulomb orbit can be extended to
conformal symmetry which is also used in the present article.

The present \textit{classical} analysis also may have implications for the
idea of thermal effects of acceleration appearing in quantum field theory.
\ The present \textit{classical} analysis involves a difference in relative
velocity $\mathbf{V}$ between the initial and final inertial frames, but does
not specify the magnitude of the acceleration. \ We can have a small
acceleration acting for a long time or a large acceleration acting for a short
time, and the change in velocity is the same. \ This suggests that the
acceleration provides a test for the stability of the distribution, but the
actual acceleration value is of little importance within classical theory.
\ The use of the accelerating Rindler frame is only to test for equilibrium in
the inertial frame. \ Thus our requirement is that an equilibrium distribution
in an inertial frame should also be an equilibrium distribution at any spatial
location within a Rindler frame. \ 

\section{Outline of the Article}

In the initial sections, we briefly discuss conformal symmetry and also a
Rindler coordinate frame acting for a finite time interval between two
inertial frames. The invariance of the average values despite Rindler
accelerations provides a test of equilibrium for random distributions. \ We
then point out that, based upon our suggestion, we can obtain the J\"{u}ttner
distribution for relativistic free \textit{particles}. \ In the analysis for
relativistic \textit{waves, }we temporarily change to two spacetime dimensions
in order to note that equilibrium in both an inertial frame and a Rindler
frame requires an infinite number of relativistic waves. \ Thermal equilibrium
requires the introduction of classical zero-point radiation whose two-point
correlation function depends upon the geodesic separation of the spacetime
points where it is evaluated. \ We note two separate types of time which enter
the Rindler coordinate frame, and we derive the full Planck spectrum with
zero-point radiation within classical theory. \ Finally, we mention that the
classical point of view does not give any support for the thermal effects of
acceleration appearing in quantum field theory. \ 

\section{Preliminary Remarks on Conformal Symmetry}

\subsection{Units and Conformal Symmetry}

Classical electromagnetism can incorporate the three fundamental
electromagnetic constants found in nature: the elementary charge $e$, the
speed of light in vacuum $c$, and the scale of classical zero-point radiation
$\hbar$. \ Therefore, there is one parameter for an electromagnetic system
which is determined by the choice of units. \ For example, the choice of a
mass $M$ (such as choosing the electron) as giving \textit{unit} scale will
then determine all quantities providing unit length, unit time, and unit
energy as
\begin{equation}
l=\frac{e^{2}}{Mc^{2}},\text{ \ \ }t=\frac{e^{2}}{Mc^{3}},\text{ \ \ }%
U=Mc^{2}\text{. \ }%
\end{equation}
The fine structure constant $e^{2}/\left(  \hbar c\right)  $ is the one
constant in classical electrodynamics which is independent of the choice of
units involving the choice of a unit mass $M$. \ If one \textit{changes} the
unit mass from $M$ over to $M^{\prime}=M/\sigma$, then all lengths change to
$l^{\prime}=\sigma l$, all times change to $t^{\prime}=\sigma t$, and all
energies change to $U^{\prime}=U/\sigma$. \ We say that the system undergoes a
$\sigma_{ltU^{-1}}$-dilation or rescaling. \ 

\subsection{Conformal Symmetry in Classical Electrodynamics}

Conformal symmetry is the group of transformations which includes those of the
Poincar$\acute{e}$ group together with the continuous, local changes of scale
corresponding to changing the choice of unit mass $M$.\cite{Kastrup} \ All of
classical electrodynamics is conformal invariant. \ Thus, 1) Maxwell's
equations are conformal covariant.\cite{C-B} \ 2) Newton's second law
$d\mathbf{p/}dt=q\mathbf{E+}q\left(  \mathbf{v/}c\right)  \times\mathbf{B}$
with forces given by electromagnetic fields $\mathbf{E}$ and $\mathbf{B}$ is
conformal covariant, since $d\mathbf{p/}dt$ scales as $1/\sigma^{2}$ as do the
fields. \ 3) Classical electromagnetic zero-point radiation is conformal
invariant.\cite{B1989} \ In this article, we will consider uniform
$\sigma_{ltU^{-1}}$-dilations of the conformal group. \ If classical
electromagnetism were the only theory in nature, the theory would be invariant
under a $\sigma_{ltU^{-1}}$-dilation since all ratios would remain unchanged. \ 

\subsection{Conformal Symmetry for Relativistic Particles and Waves}

Within the present analysis, conformal symmetry is used in an essential way
for both relativistic particles and relativistic waves. \ A homogeneous wave
equations has exactly one parameter, the speed $c$\ of the relativistic waves.
\ It follows that the speed of the wave is unchanged under Lorentz
transformation or under conformal transformation. \ Thus, a homogeneous
\textit{wave} equation is invariant under a conformal transformation. \ 

\section{Acceleration in a Rindler Frame for a Finite Time}

\subsection{A Rindler Frame Between Inertial Frames}

In a relativistic theory, the acceleration between two relativistic inertial
frames with relative velocity $\mathbf{V}$ is naturally provided by a Rindler
frame acting for a finite time interval.$~$ \ The Rindler frame involves
accelerations but no extraneous variables when changing inertial frames.

\subsubsection{Connections Between the Inertial Frames and the Rindler
Frame\ \ }

We choose the coordinates of our analysis so that the relative velocity
$\mathbf{V}$ between the initial inertial frame $S_{i}$ and the final inertial
frame $S_{f}$ is along the $x$-axis, $\mathbf{V=}\widehat{x}V$. \ If the
initial relativistic inertial frame $S_{i}$ has coordinates $\left(
ct,x,y,z\right)  ,$then the Rindler frame $R$ can have coordinates $(\eta
,\xi,y,z)$ where\cite{Schutz}
\begin{equation}
ct=\xi\sinh\eta\text{ \ and \ \ }x=\xi\cosh\eta.\text{ } \label{Rindlera}%
\end{equation}
The acceleration at any point $(\eta,\xi,y,z)$ of the Rindler frame $R$ is a
constant proper acceleration $a\left(  \xi\right)  $ given by
\begin{equation}
a\left(  \xi\right)  =\frac{c^{2}}{\xi}, \label{accx}%
\end{equation}
and so varies with the distance $\xi$ from the event horizon at $\xi=0$. \ The
varying acceleration $a\left(  \xi\right)  =c^{2}/\xi$ with distance from the
event horizon is required by relativity and is inescapable. \ If the
acceleration is made shorter, the time interval to go from $S_{i}$ to $S_{f}$
is necessarily longer. \ The variation of acceleration with height $\xi$ is
associated with the relativistic lack of synchronization for clocks in two
different \textit{relativistic} inertial frames with relative velocity
$\mathbf{V}$ between the inertial frames. \ In a Rindler coordinate frame, one
may change one's acceleration $a\left(  \xi\right)  $ simply by moving to a
new coordinate location with a different $\xi$.

If the acceleration is applied at the initial time $ct=0=\eta_{i}$, then, at
this time, the coordinates $(\eta,\xi,y,z)$ of the Rindler frame $R$ agree
exactly with the coordinates $\left(  ct,x,y,z\right)  $ of the Minkowski
inertial frame $S_{i}$, and the first-order in time changes also agree. \ If
the acceleration ceases at the Rindler coordinate time $\eta_{f}$, then the
coordinates $(\eta_{f},\xi,y,z)$ of the Rindler frame $R$ can agree with the
coordinates $\left(  ct^{\prime},x^{\prime},y,z\right)  $ of the Minkowski
inertial frame $S_{f}$ where
\begin{equation}
ct^{\prime}=\xi\sinh\eta_{f}=\gamma\left(  ct-Vx/c\right)  , \label{tpxh}%
\end{equation}
while, for fixed coordinate $\xi,$ $d\left(  ct^{\prime}\right)  =\xi\left[
\cosh\eta\right]  d\eta$, and%
\begin{equation}
x^{\prime}=\xi\cosh\eta_{f}=\gamma\left(  x-Vt\right)  , \label{xpx}%
\end{equation}
with $\gamma=\left[  1-\left(  V/c\right)  ^{2}\right]  ^{-1/2}=\cosh\eta_{f}%
$. \ The relative speed between the Minkowski frames $S_{i}$ and $S_{f}$ is
\begin{equation}
V=c\tanh\eta_{f}. \label{Vceta}%
\end{equation}
We notice that this velocity $\mathbf{V}$ is unchanged under any
$\sigma_{ltU^{-1}}$-dilation of the conformal group.

\subsubsection{Different Equations of Motion in $S_{i}$ and $R$}

At time $t=\eta=0$, both the inertial frame $S_{i}$ and the Rindler frame $R$
agree on the initial conditions of the system. \ These initial condition
involve the position and velocities of any particles, or the values of the
electromagnetic fields $\mathbf{E}$ and $\mathbf{B}$, or the values of a
scalar field $\phi$ and its first time derivative. \ The initial conditions
involve \textit{first} time derivatives and may be the same, but the
\textit{second} time derivatives are required for the equations of motion
involving Newton's 2nd law for the particles, or the wave equation for
the\ fields $\mathbf{E}$ and $\mathbf{B}$ and $\phi$. \ And the acceleration
is reflected only in the \textit{second} time derivative, which is different
for $S_{i}$ and for $R$. \ The particle trajectories or the wave normal modes
are \textit{different} in the inertial frame $S_{i}$ and in the Rindler frame
$R$. \ 

If a random distribution of particles or of waves is in equilibrium, then it
can be moved by a Rindler coordinate frame $R$ from one inertial frame $S_{i}$
to $S_{f}$ while retaining all its fundamental aspects, including its
temperature $T$. \ However, if the system is \textit{not} in equilibrium, the
fundamental averages of the system may be quite different in $S_{i}$ and in
$S_{f}$. \ For example, one free particle bouncing elastically in a box may
have a different energy in $S_{i}$ and in $S_{f}~$depending on the spatial
location of the particle in the box when the acceleration started and ended.
\ The spatial location corresponds to the phase of the particle motion or of
the wave oscillation. \ However, if the velocity distribution for many
particles is in an equilibrium situation allowed by both $S_{i}$ and $R$ at
time $t=\eta=0$, then at any coordinate time $\eta_{f}$, we will find that $R$
gives an arrangement consistent with equilibrium in $S_{f}$. \ 

\subsection{Conformal Symmetry and Time in a Rindler Frame}

\subsubsection{Acceleration in a Rindler Frame}

The spatial coordinates $\left(  \xi,y,z\right)  ~$of a point at rest in the
Rindler frame are not changing in time. \ The situation is analogous to living
in a tall apartment house. \ Any item supported by the floor is not
accelerating relative to the floor. \ However, if we drop an object out a
window, the object falls with an acceleration. \ It is this acceleration
$a\left(  \xi\right)  $ which we refer to in a Rindler frame. \ The
acceleration is that measured in the momentarily comoving inertial frame
$S^{\prime}$ with coordinates $\left(  ct^{\prime},x^{\prime},y,z\right)  $
with respect to $\left(  \xi,y,z\right)  $ where $x^{\prime}\left(  t^{\prime
}\right)  =\xi$. \ \ 

At the initial time, the acceleration $a\left(  \xi\right)  =c^{2}/\xi$ in Eq.
(\ref{accx}) comes (while holding $\xi$ constant) from the second derivative
of the coordinate $x\left(  t\right)  =\xi$ with respect to the time $t~$at
time $ct=\eta=0$ using the proper time interval arising from a unit given by a
mass $M$. \ Using the connections given in Eq. (\ref{Rindlera}), we find that
the inertial frame $S_{i}$ sees the coordinate point $\left(  \xi,y,z\right)
$ as given by $x=\sqrt{\xi^{2}+c^{2}t^{2}},$ with $y$ and $z$ unchanged.
\ Then the first derivative is $\left(  dx/dt\right)  _{\xi}=c^{2}t/\sqrt
{\xi^{2}+c^{2}t^{2}}$ which vanishes at time $t=\eta=0$. \ However, the second
time derivative is $\left(  d^{2}x/dt^{2}\right)  _{\xi}=c^{2}/\sqrt{\xi
^{2}+c^{2}t^{2}},$which is non-vanishing at the start time $t=0$, but rather
gives $a\left(  \xi\right)  =c^{2}/\xi$ which appears in Eq. (\ref{accx}). \ 

\subsubsection{Relativistic Variations in Time}

The situation is just like dropping an object out a window, except for the
relativity. \ Although the acceleration at a fixed coordinate point $\left(
\xi,y,z\right)  $ in the Rindler frame is constant in time at $a\left(
\xi\right)  =c^{2}/\xi$, there is a different acceleration (in relativity)
depending on the coordinate location $\xi$ relative to the event horizon.
\ This change fits with the fact that, in relativity, clocks which are
synchronized in one \textit{inertial} frame will not be synchronized in
another \textit{inertial} frame moving with a velocity $\mathbf{V}$ relative
to the first. \ Whereas time is absolute in the nonrelativistic physics of our
everyday experience, time transformations may depend on spatial location in relativity.

We can imagine each spacetime point at rest in a box supplied with two clocks,
one reading \textit{proper} time (dependent on the choice of $M$) and a second
reading \textit{coordinate} time $\eta$ in $R$. \ In the inertial frame
$S_{i}$, all the proper-time clocks are synchronized. \ Now the box with its
clocks is accelerated uniformly in the Rindler frame $R$ so as to come into
the $S_{f}$ inertial frame. \ In an accelerating Rindler frame, two clocks
reading proper time will run at different rates if the clocks are at different
heights above the event horizon at $\xi=0$. \ However, the \textit{coordinate}%
-time clocks will be synchronized in $R$. \ When the acceleration suddenly
ceases, the coordinate-time clocks reading Rindler \textit{coordinate} time
$\eta$ will be precisely synchronized in the $S_{f}$ inertial frame. \ The
\textit{proper} time is different depending upon the height $\xi$ above the
event horizon at $\xi=0$. \ If one starts at $ct=\eta=0$ for any positive
values of $\xi$, then the proper time read on a clock at height $\xi$ will be
out of synchronization in $S_{f}$. \ 

\subsubsection{Conformal Invariance of the Rindler Frame}

If a $\sigma_{ltU^{-1}}$-dilation is applied to a Minkowski frame or Rindler
frame where there are only point masses, one would never know a dilation had
occurred. \ The spacetime is unchanged. \ To be sure, the acceleration of a
specific coordinate point $\left(  \xi,y,z\right)  $ changes under conformal
dilation (different choice of $M$), but the distance to the event horizon also
changes, leaving the acceleration equation $a\left(  \xi\right)  =c^{2}/\xi$
in (\ref{accx}) unchanged. \ The coordinate time $\eta$ is unchanged by a
$\sigma_{ltU^{-1}}$-dilation.. \ A relativistic particle following a
trajectory given by Newton's second low would still be following such a
trajectory. \ A standing relativistic wave remains such in the Rindler frame. \ 

\section{Thermal Effects of Acceleration: Relativistic Particles}

\subsection{Relativistic Particles in the Rindler Frame}

\subsubsection{Small Acceleration Limit}

Traditional classical statistical mechanics is usually concerned with massive
particles in a \textit{nonrelativistic} context. \ Indeed, the
\textit{no-interaction theorem} for \textit{relativistic} \textit{mechanics}
allows only \textit{point collisions} between relativistic point
particles.\cite{NOINT} \ If any position-dependent potential function is
introduced, one must go to a full field theory. \ On this account, Boltzmann
statistical mechanics treats \textit{nonrelativistic} particles and
potentials. \ 

Here we consider \textit{relativistic free} particles, and find the
J\"{u}ttner distribution for free, noninteracting, relativistic particles of
rest mass $m_{0}$. \ Since the acceleration of the Rindler frame is arbitrary,
we will consider a box where the end closest to the event horizon is at
$\xi_{0}$ with \textit{small} acceleration, $a\left(  \xi_{0}\right)
=c^{2}/\xi_{0}$, and the acceleration at the other end of the box at $\xi
_{0}+L$ is even smaller. \ In this small-acceleration region, we may take the
height $L$ small compared to the distance $\xi_{0}$ to the event horizon,
$L/\xi_{0}<<1,$ and have $\left[  a\left(  \xi\right)  \right]  \xi/c^{2}<<1$.
\ Accordingly, we may use the small-gravitational-field approximation, and may
use nonrelativistic physics for the potential energy function with a constant
acceleration for the box. \ Therefore, for particles with relativistic kinetic
energy, we are dealing with an (approximate) total energy
\begin{equation}
m_{0}c^{2}\left(  \gamma-1\right)  +m_{0}\left[  a\left(  \xi\right)  \right]
\xi=m_{0}c^{2}\left\{  \gamma-1+\left[  a\left(  \xi\right)  \right]
\xi/c^{2}\right\}  \approxeq m_{0}c^{2}\left[  \gamma-1+a\left(  \xi
_{0}\right)  \xi/c^{2}\right]  .
\end{equation}
Note the change from $\xi$ over to $\xi_{0}$. \ Thus we may use the
relativistic expression $m_{0}c^{2}\left(  \gamma-1\right)  $ for the kinetic
energy and the nonrelativistic expression $m_{0}a\xi=m_{0}\left[  a\left(
\xi_{0}\right)  \right]  \xi$ for the potential energy. \ 

\subsubsection{Barometric Equation}

A statistical mechanical analysis requires the barometric equation\cite{Boltz}
for the spatial distribution of heights $\xi$ in the Rindler frame, since only
rest mass is contributing to the potential energy function. \ Therefore just
as in the nonrelativistic case, the kinetic energy distribution is constant
with height, though here the kinetic energy involves the relativistic
expression $m_{0}c^{2}\left(  \gamma-1\right)  $. \ %

\begin{equation}
\mathcal{P}_{R}\left(  \gamma,m_{0},a\left(  \xi_{0}\right)  ,L\right)
=const\times\exp\left[  -\frac{m_{0}\left(  \gamma-1\right)  }{k_{B}T}\right]
\exp\left[  -\frac{m_{0}a\left(  \xi_{0}\right)  \xi}{k_{B}T}\right]  .
\label{PRMT}%
\end{equation}

Now at the initial time when the box is suddenly accelerated in the Rindler
frame, the pure kinetic energy $m_{0}c^{2}\left(  \gamma_{I}-1\right)  $ of a
particle in the inertial frame $S_{i}$ suddenly becomes total energy
(including gravitational potential energy) $m_{0}c^{2}\left[  \gamma
_{I}-1+a\left(  \xi_{0}\right)  \xi_{I}/c^{2}\right]  $ in the Rindler frame
$R$. \ This total energy is maintained constant in the Rindler frame $R$ until
the acceleration stops. Then the final kinetic energy $m_{0}c^{2}\left(
\gamma_{F}-1\right)  $ of the particle in the $S_{f}$ inertial frame equals
the kinetic energy in the accelerating Rindler frame $R$. \ Equilibrium in the
inertial frames $S_{i}$ and $S_{f}$ holds if the distribution of the particle
speeds is unchanged (equilibrium holds), only if on average $\left\langle
\gamma_{F}\right\rangle =\left\langle \gamma_{I}\right\rangle $ despite
acceleration in the Rindler frame. \ But in the Rindler frame in equilibrium,
the kinetic energy distribution matched the barometric equation. \ Thus we
have%
\begin{equation}
\mathcal{P}_{S}\left(  \gamma,m_{0},L\right)  =const\times\exp\left[
-\frac{m_{0}c^{2}\left(  \gamma-1\right)  }{k_{B}T}\right]  . \label{Jut}%
\end{equation}
A general distribution of particles would not satisfy this requirement that
$\left\langle \gamma_{F}\right\rangle =\left\langle \gamma_{I}\right\rangle $
after being accelerated in a Rindler frame.

We notice that conformal symmetry is satisfied since $m_{0}c^{2}/\left(
k_{B}T\right)  $ is constant under a $\sigma_{ltU^{-1}}$-scale change. \ The
expression in Eq. (\ref{Jut}) matches the J\"{u}ttner
expression.\cite{Juttner} \ The undetermined $const$ reflects the
normalization constants associated with the number of directions of particle
motion and the temperature normalization associated with the total energy and
the total number of particles in the box. \ 

\subsubsection{Qualitative Explanation}

A qualitative understanding of what is involved can be found in our familiar
ideas of gravity in connection with a pile of books sitting on a floor of a
building. \ First, we must be aware of the increase in supporting force needed
by lower books. \ The bottom book must provide a supporting force balancing
the weight of all those books above it. \ Second, for \textit{free particles,}
whether nonrelativistic or relativistic, the pressure at any point is provided
by the kinetic energy of any moving particles. \ Third, if the floor support
under the pile of books were suddenly removed, the books would fall freely
toward the basement. \ While the books are falling freely, they are in an
inertial frame and all have the same velocity. \ Just before the bottom book
hits the basement floor, the momentarily comoving inertial frame at rest with
respect to the basement floor is our inertial frame $S_{i}$ while the building
provides the Rindler frame $R$. \ At this moment, all the books have the same
velocity in both frames. \ However, the bottom book falls a smaller distance
to reach the basement floor than the top book which would travel further, and
hence be going faster if it hit the basement floor. \ Thus although all the
books initially had zero speed relative to the basement floor, they would have
different speeds after they hit the basement floor since some books fall
further than others in the gravitational field. \ These qualitative ideas
provide a classical understanding for equilibrium in an inertial frame when
the system is carried from one inertial frame to another by a Rindler frame. \ 

\subsubsection{Importance of Phase in the Periodic Motions}

The \textit{initial }conditions involving particle positions and velocities
are identical between $S_{i}$ and $R$, while the equations of motion in the
box are different between $S_{i}$ and $R$. \ The actual motions are determined
by the initial conditions and the equations of motion. \ The average energy is
determined by the initial conditions and does not \textit{change} in either
coordinate frame. \ In general, the energy of the particles will be different
in $S_{i}$ and in $R$. \ If the random system is in equilibrium in both
$S_{i}$ and in $R$, then the system is truly in equilibrium. \ The initial
\textit{kinetic energy} depends only on the particle velocities and is the
same in both $S_{i}$ and $R$, but the inclusion of gravitational potential
energy in the Rindler frame makes for a difference. \ Thus, for an
\textit{equilibrium} situation, the Rindler frame delivers the system to the
final inertial frame $S_{f}$ with the same average kinetic energy as in the
initial inertial frame $S$.

\subsubsection{Conformal Invariance of the Probability}

We emphasize that in both the Rindler frame and the inertial frame, the
probabilities in Eqs. (\ref{PRMT})\ and (\ref{Jut}) are invariant under the
$\sigma_{ltU^{-1}}$-scaling of conformal transformation. \ 

\section{ Thermal Effects of Acceleration: Two-Dimensional Relativistic
Waves\ }

\subsection{Waves in a Box}

\subsubsection{Inertial Frame Radiation for Two-Dimensional Scalar Waves in a
Box}

We would like to repeat our argument corresponding to accelerating a box in
thermal equilibrium, but now, instead of relativistic massive particles, we
consider relativistic waves. \ In this case, it is the \textit{phase} of the
waves which is all-important. \ Classical \textit{relativistic} waves travel
at the speed of light $c$. \ 

In order to simplify the situation as much as possible, we consider
\textit{scalar} relativistic waves $\phi(ct,x)$ in only \textit{two spacetime
dimensions} in the inertial frame $S$ with a Lagrangian density\cite{2D}%
\begin{equation}
\mathcal{L=}\frac{1}{2}\left[  \left(  \frac{1}{c}\frac{\partial\phi}{\partial
t}\right)  ^{2}-\left(  \frac{\partial\phi}{\partial x}\right)  ^{2}\right]  ,
\label{Lagr}%
\end{equation}
and with Hamiltonian
\begin{equation}
H=\frac{1}{2}\int dx\left[  \left(  \frac{\partial\phi}{\partial x}\right)
^{2}+\left(  \frac{1}{c}\frac{\partial\phi}{\partial t}\right)  ^{2}\right]  .
\label{Ham}%
\end{equation}
In the inertial frame $S$, the waves satisfy the wave equation
\begin{equation}
\left(  \frac{\partial^{2}\phi}{\partial x^{2}}\right)  -\frac{1}{c^{2}%
}\left(  \frac{\partial^{2}\phi}{\partial t^{2}}\right)  =0. \label{waveq}%
\end{equation}
In an inertial frame, the solutions of this wave equation, together with
Dirichlet boundary conditions in a one spatial dimensional box of length $L$
oriented along the $x$-axis, are standing waves of the form%
\begin{equation}
\phi_{n}(ct,x)=A_{n}\sqrt{\frac{2}{L}}\sin\left[  k_{n}x\right]  \cos\left[
\omega_{n}t-\theta(k_{n})\right]  , \label{wavew}%
\end{equation}
where $k_{n}=n\pi/L,$ $n=1,2,..$, the angular frequency is $\omega_{n}=ck_{n}%
$, and the phase $\theta(k_{n})$ is random on $[0,2\pi)$ and independent for
each $k_{n}$ corresponding to random radiation.\cite{Rice} \ The connection
between the amplitude $A_{n}$ and the energy $U_{n}$ per normal mode of the
wave in the container is%

\begin{equation}
U_{n}=\frac{1}{2}\left[  \frac{\omega_{n}}{c}A_{n}\right]  ^{2}. \label{Uw}%
\end{equation}
\ 

\subsubsection{Rindler Frame Radiation Normal Modes}

In the accelerating Rindler frame $R$, the differential equation for
$\phi\left(  ct,x\right)  =\phi\left(  \xi\sinh\eta,\xi\cosh\eta\right)
=\varphi\left(  \eta,\xi\right)  $ satisfies not the wave equation in
(\ref{waveq}) but rather the equation%
\begin{equation}
\frac{1}{\xi}\frac{\partial}{\partial\xi}\left(  \xi\frac{\partial\varphi
}{\partial\xi}\right)  -\frac{1}{\xi^{2}}\left(  \frac{\partial^{2}\varphi
}{\partial\eta^{2}}\right)  =0. \label{Rindlereq}%
\end{equation}
We obtain this equation from Eq. (\ref{waveq}) by the usual rules of partial
differentiation from Eq. (\ref{Rindlera}). \ It satisfies the correct
$\sigma_{ltU^{-1}}$-dilation behavior where $\xi$ is a length, and $\eta$ is
invariant under $\sigma_{ltU^{-1}}$-dilation.

If one considers normal modes of oscillation in the Rindler frame, one finds
an equation (\ref{Rindlereq}) of the Sturm-Liouville form with a weight factor
$1/\xi$ and an eigenvalue $\beta_{n}^{2}$ where%
\begin{equation}
\beta_{n}=\frac{n\pi}{\ln\left[  \left(  \xi_{0}+L\right)  /\xi_{0}\right]  }.
\label{betan}%
\end{equation}
Here $\beta_{\mathbf{n}}$ (like the Rindler coordinate time $\eta$) is
conformal covariant. \ A normal mode of oscillation in $R$ takes the form
\begin{equation}
\varphi\,_{n}(\eta,\xi)=\mathcal{A}_{n}\left\{  \sqrt{\frac{2}{\ln\left[
\left(  \xi_{0}+L\right)  /\xi_{0}\right]  }}\sin\left[  \frac{n\pi}%
{\ln\left[  \left(  \xi_{0}+L\right)  /\xi_{0}\right]  }\ln\left[  \frac{\xi
}{\xi_{0}}\right]  \right]  \right\}  \cos\left[  \beta_{n}\eta-\theta\left(
\mathbf{k}_{n}\right)  \right]  , \label{Rindsol}%
\end{equation}
We notice that each mode vanishes at the ends $\xi_{0}$ and $\xi_{0}+L$ of the
container, since $\ln(\xi_{0}/\xi_{0})=0$, while $\ln\left[  (\xi_{0}%
+L)/\xi_{0}\right]  /\ln\left[  (\xi_{0}+L)/\xi_{0}\right]  =1,$ and
$\sin\left[  n\pi\right]  =0\,\ $for integer $n.$ \ The Rindler normal mode
$\varphi_{n}(\eta,\xi)$ in Eq. (\ref{Rindsol}) is not at all the same as the
inertial frame mode $\phi_{n}\left(  ct,x\right)  $ suggested in Eq.
(\ref{wavew}).

The energy of this mode at time $t=\eta=0$ follows from the Hamiltonian in Eq.
(\ref{Ham}) as
\begin{align}
U_{n}  &  =\frac{1}{2}\int_{\xi_{0}}^{\xi_{0}+L}\frac{d\xi}{\xi}\,\left[
\mathcal{A}_{n}\right]  ^{2}\frac{2}{\ln\left[  \left(  \xi_{0}+L\right)
/\xi_{0}\right]  }\sin^{2}\left[  \frac{n\pi}{\ln\left[  \left(  \xi
_{0}+L\right)  /\xi_{0}\right]  }\ln\left[  \frac{\xi}{\xi_{0}}\right]
\right] \nonumber\\
&  \times\left\{  \left(  \frac{n\pi}{\ln\left[  \left(  \xi_{0}+L\right)
/\xi_{0}\right]  }\right)  ^{2}\cos^{2}\left[  \beta_{n}\eta-\theta\left(
k_{n}\right)  \right]  +\beta_{n}^{2}\sin^{2}\left[  \beta_{n}\eta
-\theta\left(  k_{n}\right)  \right]  \right\} \nonumber\\
&  =\frac{1}{2}\int_{0}^{1}du\left(  2\beta_{n}^{2}\right)  \left[
\mathcal{A}_{n}\right]  ^{2}\sin^{2}\left[  n\pi u\right]  =\frac{1}{2}%
\beta_{n}^{2}\left[  \mathcal{A}_{n}\right]  ^{2}, \label{UA}%
\end{align}
where we have used $u=\ln\left[  \xi/\xi_{0}\right]  /\ln\left[  \left(
\xi_{0}+L\right)  /\xi_{0}\right]  $ in evaluating the integral. \ Note that
$\beta_{n}^{2}$ in Eqs. (\ref{betan}) and (\ref{UA}) is not the same as
$k_{n}=n\pi/L$.

\subsubsection{Different Normal Modes of Oscillation}

The \textit{initial conditions} are the same in $S_{i}$ and $R$, but the
\textit{normal modes of oscillation }given are quite different between the
inertial frame $S_{i}$ and the Rindler frame $R$. \ Indeed, although the
solutions given in Eqs. (\ref{wavew}) and (\ref{Rindsol}) both involve
integers $n$, \textit{a normal mode corresponding to a single integer }%
$n$\textit{ in one frame will have nonzero expansion coefficients involving an
infinite number of integers }$n$\textit{ in the other frame. \ }

\subsubsection{Equilibrium in the Rindler Frame}

Only very special radiation spectra will give equilibrium under acceleration
for relativistic waves. In the previous example involving massive particles,
we were able to pick out the familiar barometric equation at constant
temperature as giving equilibrium in the Rindler frame $R$. \ Here we would
like to find an analogous situation for relativistic waves. \ \ 

The crucial equilibrium aspect for waves involves \textit{relative phase} in
the oscillation, just as it was for particles where it was the \textit{phase}
in the particle's periodic trajectory. \ Waves of different frequencies
interfere, and the interference between the normal modes in $S_{i}$ and in
$R\,$\ will be quite different. \ Indeed, any distribution of a
\textit{finite} number of waves in the inertial frame $S_{i}$ will lead to an
\textit{infinite} number of normal modes in $R$, which will then lead to an
\textit{infinite} number of normal modes in $S_{f}$. \ \textit{Under
scattering by acceleration, an infinite number of normal modes is unavoidable
for classical radiation equilibrium involving relativistic waves. \ }

\subsection{Classical Zero-Point Radiation}

\subsubsection{Needed Idea}

Having now pointed out that a Rindler frame tends to mix \textit{wave} modes
and suggests an \textit{infinite number of modes}, we go back to four
spacetime dimensions for the ease of the subsequent analysis. \ 

\textit{Nonrelativistic} classical physics always involves a finite number of
particles in a finite-volume box, and accordingly there are always a
\textit{finite} number of \textit{particle-based} waves in the box. \ However,
there are an \textit{infinite} number of \textit{relativistic} waves in a
finite-volume box with conducting walls. \ Therefore, if there is a
\textit{finite} amount of thermal energy in the box, the energy cannot be
assigned equally to all the \textit{infinite} number of relativistic wave
modes. \ There must be some \textit{relativistic} idea which confines the
thermal radiation to the lower frequency modes in the box. \ 

The needed classical idea involves classical electromagnetic zero-point
radiation. \ Classical physics allows source-free fields $\mathbf{E}$ and
$\mathbf{B}$ or $\phi$ which satisfy the \textit{conformal-invariant wave
equation}, but which have \textit{no sources}. \ \textit{Classical}
\textit{zero-point radiation is the unique equilibrium spectrum of random
radiation which is isotropic in every inertial frame. \ }Classical
electromagnetic zero-point radiation is Lorentz-invariant, $\sigma_{ltU^{-1}}%
$-scale invariant, conformal invariant, and has the same appearance in every
inertial frame.\cite{zpr} \ It is the infinite-dimensional representation of
the conformal group associated with random radiation. \ Each normal mode of
oscillation is a basis vector in the representation space. \ 

\subsubsection{\ Classical Zero-Point Radiation and Casimir Forces}

The $\sigma_{ltU^{-1}}$-scaling constant for zero-point radiation can be
determined by experiments involving Casimir forces and electromagnetic
fields.\cite{Casimir}\cite{CForces} \ It is found to an accuracy of close to
1\%, that the best description of the source-free fields is a
Lorentz-invariant spectrum of energy $U\left(  \omega\right)  $ per normal
mode at frequency $\omega$ of random classical radiation with a scale given by
a constant $\hbar$,
\begin{equation}
U\left(  \omega\right)  =\frac{1}{2}\hbar\omega, \label{zpr}%
\end{equation}
where $\hbar$ agrees in numerical value with Planck's constant. \ The constant
$\hbar$ appeared first, not quantum theory, but in the classical analysis of
blackbody radiation in 1899.\cite{constant} \ The \textit{classical}
electromagnetic scale factor $\hbar$ in Eq. (\ref{zpr}) has nothing to do with
energy quanta. \ However, some physicists have never seen $\hbar$ in any
context other than quantum theory, and so are surprised to find it a
\textit{classical} theory. \ 

Since classical zero-point radiation involves an \textit{infinite} number of
radiation modes of increasing energy, the total zero-point energy per unit
volume is divergent. \ This divergence is irrelevant since the spectrum leads
to finite Casimir \textit{forces}. \ The electromagnetic analogy to this
divergence corresponds to an infinite sheet of uniform charge density which
leads to an electric field; the total charge is divergent, but the
electromagnetic fields and forces are finite. \ 

\subsubsection{Geodesic Connection for the Correlation Function}

Zero-point radiation is invariant under the $\sigma_{ltU^{-1}}$-dilation of
conformal symmetry. \ Thus the \textit{spectrum} of classical zero-point
radiation in Eq. (\ref{zpr})\ corresponds to an energy per normal mode at
frequency $\omega$ in an inertial frame. \ This is the only connection between
energy and frequency which satisfies conformal $\sigma_{ltU^{-1}}$-scaling,
since both $U$ and $\omega$ transform as $1/\sigma$ while $\hbar$ is an
invariant. \ \ 

In the subsequent analysis for four spacetime dimensions, we will consider the
correlation function for zero-point radiation $\left\langle \phi
(ct,x,y,z)\phi\left(  ct^{\prime},x^{\prime},y^{\prime},x^{\prime}\right)
\right\rangle $ between the two spacetime coordinate points. \ The correlation
function itself must transform under conformal $\sigma_{ltU^{-1}}$-scaling as
$1/\sigma^{2}$ for four spacetime dimensions. \ However, zero-point radiation
is conformal invariant and has no quantities giving a preferred length or time
or energy. \ Therefore the correlation function must depend upon only the
geodesic separation $s^{2}=\left(  ct-ct^{\prime}\right)  ^{2}-\left(
x-x^{\prime}\right)  ^{2}-\left(  y-y^{\prime}\right)  ^{2}-\left(
z-z^{\prime}\right)  ^{2}$ between the spacetime coordinates $(ct,x,y,z)$ and
$\left(  ct^{\prime},x^{\prime},y^{\prime},x^{\prime}\right)  $. \ For
infinite space, the only such correlation function with the correct
$\sigma_{ltU^{-1}}$-dilation under conformal symmetry is
\begin{equation}
\left\langle \phi(ct,x,y,z)\phi\left(  ct^{\prime},x^{\prime},y^{\prime
},x^{\prime}\right)  \right\rangle =\frac{const}{\left(  ct-ct^{\prime
}\right)  ^{2}-\left(  x-x^{\prime}\right)  ^{2}-\left(  y-y^{\prime}\right)
^{2}-\left(  z-z^{\prime}\right)  ^{2}}, \label{coorphi}%
\end{equation}
where the constant can depend upon $\hbar$. \ 

For a box of finite spatial volume $L^{3}$, the scalar field for zero-point
radiation in an inertial frame could be written as
\begin{align}
&  \phi(ct,x,y,z)\nonumber\\
&  =\sum\nolimits_{\mathbf{k}_{\mathbf{n}}}\sqrt{\frac{4\pi U\left(
ck\right)  }{L^{3}ck^{2}}}\sqrt{\frac{2}{L}}\sin\left[  \frac{n_{x}\pi}%
{L}x\right]  \sqrt{\frac{2}{L}}\sin\left[  \frac{n_{y}\pi}{L}y\right]
\sqrt{\frac{2}{L}}\sin\left[  \frac{n_{z}\pi}{L}z\right]  \cos\left[
ck_{\mathbf{n}}t-\theta(\mathbf{k}_{\mathbf{n}})\right]  ,\label{phiUct}%
\end{align}
where we have used the expressions in Eqs. (\ref{wavew}) and (\ref{Uw}) for
four spacetime dimensions. \ Then it follows that%

\begin{align}
&  \left\langle \phi(ct,x,y,z)\phi\left(  ct^{\prime},x^{\prime},y^{\prime
},x^{\prime}\right)  \right\rangle \nonumber\\
&  =\left\langle \sum\nolimits_{\mathbf{k}}\sqrt{\frac{2U\left(  ck\right)
}{k^{2}}}\sqrt{\frac{2}{L}}\sin\left[  \frac{n_{x}\pi}{L}x\right]  \sqrt
{\frac{2}{L}}\sin\left[  \frac{n_{y}\pi}{L}y\right]  \sqrt{\frac{2}{L}}%
\sin\left[  \frac{n_{z}\pi}{L}z\right]  \cos\left[  ck_{\mathbf{n}}%
t-\theta(\mathbf{k}_{\mathbf{n}})\right]  \right. \nonumber\\
&  \times\left.  \sum\nolimits_{\mathbf{k}^{\prime}}\sqrt{\frac{2U\left(
ck^{\prime}\right)  }{k^{2}}}\sqrt{\frac{2}{L}}\sin\left[  \frac{n_{x}%
^{\prime}\pi}{L}x^{\prime}\right]  \sqrt{\frac{2}{L}}\sin\left[  \frac
{n_{y}^{\prime}\pi}{L}y^{\prime}\right]  \sqrt{\frac{2}{L}}\sin\left[
\frac{n_{z}^{\prime}\pi}{L}z^{\prime}\right]  \cos\left[  ck_{\mathbf{n}%
}^{\prime}t^{\prime}-\theta(\mathbf{k}_{\mathbf{n}}^{\prime})\right]
\right\rangle \nonumber\\
&  =\sum\nolimits_{\mathbf{k}}\frac{2U\left(  ck\right)  }{k^{2}}\left(
\sqrt{\frac{2}{L}}\sin\left[  \frac{n_{x}\pi}{L}x\right]  \sqrt{\frac{2}{L}%
}\sin\left[  \frac{n_{x}\pi}{L}x^{\prime}\right]  \right)  \left(  \sqrt
{\frac{2}{L}}\sin\left[  \frac{n_{y}\pi}{L}y\right]  \sqrt{\frac{2}{L}}%
\sin\left[  \frac{n_{y}\pi}{L}y^{\prime}\right]  \right) \nonumber\\
&  \times\left(  \sqrt{\frac{2}{L}}\sin\left[  \frac{n_{z}\pi}{L}z\right]
\sqrt{\frac{2}{L}}\sin\left[  \frac{n_{z}\pi}{L}z^{\prime}\right]  \right)
\cos\left[  ck_{\mathbf{n}}\left(  t-t^{\prime}\right)  \right]  \frac{1}{2},
\label{zpcorr}%
\end{align}
using the averages%
\begin{equation}
\left\langle \cos\left[  \theta\left(  \mathbf{k}\right)  \right]  \cos\left[
\theta\left(  \mathbf{k}^{\prime}\right)  \right]  \right\rangle =\left\langle
\sin\left[  \theta\left(  \mathbf{k}\right)  \right]  \sin\left[
\theta\left(  \mathbf{k}^{\prime}\right)  \right]  \right\rangle =\left(
1/2\right)  \delta_{\mathbf{k}_{\mathbf{n}}\mathbf{k}_{\mathbf{n}}^{\prime}%
}^{3}%
\end{equation}
and
\begin{equation}
\left\langle \cos\left[  \theta\left(  \mathbf{k}\right)  \right]  \sin\left[
\theta\left(  \mathbf{k}^{\prime}\right)  \right]  \right\rangle =0.
\end{equation}

If we choose the spatial coordinates the same, then the correlation function
becomes for a large box (where we have replaced $\sin^{2}\left[  n\pi
x/L\right]  $ by its average value 1/2)%
\begin{equation}
\left\langle \phi^{zp}(ct,x,y,z)\phi^{zp}\left(  ct^{\prime},x,y,x\right)
\right\rangle =\sum\nolimits_{\mathbf{k}}\frac{4\pi U\left(  ck\right)
}{L^{3}k^{2}}\cos\left[  k_{\mathbf{n}}\left(  ct-ct^{\prime}\right)  \right]
.
\end{equation}
Then for a large box, we approximate the sum by an integral as $\sum
\nolimits_{\mathbf{k}}\approxeq\int_{0}^{\infty}dn_{x}\int_{0}^{\infty}%
dn_{y}\int_{0}^{\infty}dn_{z}=\left(  L^{3}/\pi^{3}\right)  \int_{0}^{\infty
}dk_{x}\int_{0}^{\infty}dk_{y}\int_{0}^{\infty}dk_{z}$ and integrate over all
solid angles to obtain%
\begin{align}
&  \left\langle \phi^{zp}(ct,x,y,z)\phi^{zp}\left(  ct^{\prime},x^{\prime
},y^{\prime},x^{\prime}\right)  \right\rangle \nonumber\\
&  =\frac{2}{\pi c}\int_{0}^{\infty}d\left(  ck\right)  U\left(  ck\right)
\cos\left[  k\left(  ct-ct^{\prime}\right)  \right]  .
\end{align}

For the case of zero-point radiation, this becomes%

\begin{align}
\left\langle \phi^{zp}(ct,x,y,z)\phi^{zp}\left(  ct^{\prime},x,y,x\right)
\right\rangle  &  =\frac{2}{\pi}\int_{0}^{\infty}dkU\left(  ck\right)
\cos\left[  k\left(  ct-ct^{\prime}\right)  \right]  \nonumber\\
&  =\frac{2}{\pi}\int_{0}^{\infty}dk\left(  \frac{\hbar}{2}ck\right)
\cos\left[  k\left(  ct-ct^{\prime}\right)  \right]  \nonumber\\
&  =\left(  \frac{\hbar c}{\pi}\right)  \left(  \frac{-1}{\left(
ct-ct^{\prime}\right)  ^{2}}\right)  ,
\end{align}
where we have used the singular Fourier transform%
\begin{equation}
\int_{0}^{\infty}dkk\cos\left[  bk\right]  =\operatorname{Re}\lim
_{\lambda\rightarrow0}\int_{0}^{\infty}dkk\exp\left[  \left(  ib-\lambda
\right)  k\right]  =-b^{-2}.
\end{equation}
Thus the correlation function for the classical zero-point radiation of a
scalar wave is%
\begin{equation}
\left\langle \phi^{zp}(ct,x,y,z)\phi^{zp}\left(  ct^{\prime},x^{\prime
},y^{\prime},x^{\prime}\right)  \right\rangle =\left(  \frac{\hbar c}{\pi
}\right)  \frac{-1}{\left(  ct-ct^{\prime}\right)  ^{2}-\left(  x-x^{\prime
}\right)  ^{2}-\left(  y-y^{\prime}\right)  ^{2}-\left(  z-z^{\prime}\right)
^{2}},
\end{equation}
exactly as suggested in Eq. (\ref{coorphi}). \ The geodesic connection with
classical zero-point radiation will remain, even when the coordinates are
changed. \ 

\subsubsection{Forces on the Ends of a Conducting-Walled Box}

Casimir forces from zero-point radiation involve small separations between
objects where the important wavelengths of the zero-point radiation spectrum
are comparable to the separations between the objects and the discrete nature
of radiation modes is crucial. \ However, here we are interested in the
situation for a large box with perfectly conducting walls. \ And zero-point
radiation is assumed to be everywhere. \ The average force per unit area on a
perfectly-reflecting plane surface due to two plane waves (which are
reflected) of the same frequency but traveling initially in opposite
directions simply vanishes. \ \ However, if we go to another inertial frame
where the conducting plane has non-zero velocity, the net average force is
again zero but the frequencies of the same two approaching waves on the
opposite sides are not the same. \ The classical zero-point spectrum is the
unique radiation spectrum which is the same in all inertial frames. \ In an
accelerating frame, the average force on the plane in classical zero-point
radiation vanishes because the spectral distribution is the same in all
inertial frames. \ 

\subsection{Zero-Point Radiation is in Equilibrium in the Rindler Frame}

\subsubsection{Zero-Point Radiation in an Inertial Frame}

Here the most natural example of a distribution of relativistic waves which
will give equilibrium in the Rindler frame is classical electromagnetic
zero-point radiation. \ This spectrum of random relativistic waves takes the
same form in every inertial frame. \ It is conformal \textit{invariant}%
.\cite{B1989} \ This classical situation has already been considered several
times with varying points of view in classical physics.\cite{constant}%
\cite{B42}\cite{Bacc} \ 

In the initial inertial frame $S_{i}$, we can wait as long as needed to
determine the time-independent correlation function $\left\langle \phi
^{zp}\left(  ct,x,y,z\right)  \phi^{zp}\left(  ct^{\prime},x^{\prime
},y^{\prime},z^{\prime}\right)  \right\rangle $. \ For classical zero-point
radiation in any inertial frame, this correlation function always depends upon
the geodesic separation of the spacetime points. \ 

\subsubsection{Spectrum Remains Zero-Point Radiation in the Comoving Inertial
Frame}

The Rindler frame has a coordinate time $\eta$ which agrees with the time in
the momentarily comoving inertial frame $S^{\prime}$at the time $\eta^{\prime
}=ct^{\prime}/\xi$, and therefore the correlation function becomes
\begin{equation}
\left\langle \phi^{zp}\left(  ct^{\prime},x,y,z\right)  \phi^{zp}\left(
ct^{\prime},x^{\prime},y^{\prime},z^{\prime}\right)  \right\rangle =\left(
\frac{\hbar c}{\pi}\right)  \frac{-1}{\left[  0-\left(  x-x^{\prime}\right)
^{2}-\left(  y-y^{\prime}\right)  ^{2}-\left(  z-z^{\prime}\right)
^{2}\right]  }.
\end{equation}

We emphasize that at a single time $t^{\prime}$ in the comoving inertial
frame, the spectrum of radiation corresponds to classical zero-point
radiation; it has no preferred length or energy. \ \ At a single time, the
spatial dependence shows no acceleration-dependent behavior. \ This situation
is familiar to people living in a tall building. \ The separations between
objects in the building do not appear to change. \ In the exactly comparable
situation involving wavelengths for random zero-point radiation, the random
spatial pattern in the accelerating Rindler frame $R$ remains the same as that
of the momentarily comoving inertial frame. \ 

\subsubsection{Zero-Point Radiation in the Rindler Frame}

In the Rindler coordinate frame, this correlation function involving a time
difference becomes%
\begin{align}
&  \left\langle \phi^{zp}\left(  \xi\sinh\eta,\xi\cosh\eta,y,z\right)
\phi^{zp}\left(  \xi\sinh\eta^{\prime},\xi\cosh\eta^{\prime},y^{\prime
},z^{\prime}\right)  \right\rangle \nonumber\\
&  =\left(  \frac{\hbar c}{\pi}\right)  \frac{-1}{\left[  \left(  \xi\sinh
\eta-\xi\sinh\eta^{\prime}\right)  ^{2}-\left(  \xi\cosh\eta-\xi\cosh
\eta^{\prime}\right)  ^{2}-\left(  y-y^{\prime}\right)  ^{2}-\left(
z-z^{\prime}\right)  ^{2}\right]  }\nonumber\\
&  =\left(  \frac{\hbar c}{\pi}\right)  \frac{-1}{\left[  -2\xi^{2}-2\xi
^{2}\cosh\left(  \eta-\eta^{\prime}\right)  -\left(  y-y^{\prime}\right)
^{2}-\left(  z-z^{\prime}\right)  ^{2}\right]  }\nonumber\\
&  =\left(  \frac{\hbar c}{\pi}\right)  \frac{-1}{\left[  4\xi^{2}\sinh
^{2}\left[  \left(  \eta-\eta^{\prime}\right)  /2\right]  -\left(
y-y^{\prime}\right)  ^{2}-\left(  z-z^{\prime}\right)  ^{2}\right]
}.\label{phizpx}%
\end{align}
At \textit{coordinate} time $\eta^{\prime}$, all coordinates $\xi$ in the
comoving inertial frame $S^{\prime}$ have the same speed relative to the
initial inertial frame $S_{i}$. \ However, the \textit{proper} time interval
involved may be quite different for different heights $\xi$. \ The correlation
function of zero-point radiation in Rindler coordinates at fixed $\left(
\xi,y,z\right)  $ becomes from Eq. (\ref{phizpx})%
\begin{equation}
\left\langle \phi^{zp}\left(  \xi\sinh\eta,\xi,y,z\right)  \phi^{zp}\left(
\xi\sinh\eta^{\prime},\xi,y,z\right)  \right\rangle =\frac{\hbar c}{\pi}%
\frac{1}{\xi^{2}}\frac{-1}{4\sinh^{2}\left[  \left(  \eta^{\prime}%
-\eta\right)  /2\right]  }.\label{X}%
\end{equation}
Here we have written the expression in terms of the coordinates $\eta,\xi,y,z$
in a Rindler frame. \ Now this expression holds for classical zero-point
radiation, and there is no reference to the acceleration in the Rindler frame
nor how long is the proper time interval. \ At any time $t^{\prime}>\eta_{f},$
where $\eta_{f}$ is the time when the Rindler acceleration ceases, this
distribution will continue as zero-point radiation in the final inertial frame
$S_{f}$. \ 

\subsection{Thermal Radiation in a Rindler Frame}

\subsubsection{Thermal Radiation Above Zero-Point Radiation}

Classical zero-point radiation does not simply disappear when we discuss
thermal radiation. \ Rather we expect that classical thermal radiation
$U^{T}\left(  \omega,T\right)  $ is radiation \textit{in addition to} the
classical zero-point spectrum $U^{zp}\left(  \omega\right)  $, so that both
are present,%
\begin{equation}
U^{total}\left(  \omega,T\right)  =U^{T}\left(  \omega,T\right)
+U^{zp}\left(  \omega\right)  . \label{Utotneed}%
\end{equation}
\ If the radiation in the inertial frame $S_{i}$ includes thermal radiation,
we expect that the total radiation spectrum will continue to include thermal
radiation in the Rindler frame $R$. \ Since the radiation pattern is
stationary in the Rindler frame, we expect it to satisfy the Tolman-Ehrenfest
relation\cite{Tolman} connecting the temperature to the coordinates as
$T\sqrt{\left\vert g_{00}\right\vert }=const$. \ For our Rindler coordinates
in flat spacetime where $ds^{2}=\xi^{2}d\eta^{2}-d\xi^{2}-dy^{2}-dz^{2}$, this
corresponds to $T\sqrt{\left\vert g_{00}\right\vert }$ $=\left[  T\left(
\xi\right)  \right]  \xi=const$. \ Thus the temperature of the radiation is
varying with the spatial coordinate in the Rindler frame, being large for
small values of $\xi$ near the event horizon, and falling smoothly toward zero
at large values of $\xi$ far from the event horizon. \ This is the
\textit{same} sort of behavior as found for the acceleration $\left[  a\left(
\xi\right)  \right]  \xi=c^{2}$ using the \textit{proper} time$\ $in $R$ and
the distance $\xi$ to the event horizon. \ 

\subsubsection{Spectrum Remains Thermal Radiation in the Comoving Inertial
Frame\ }

In the momentarily comoving inertial frame $S^{\prime}$, where the time
coordinate $t^{\prime}$ agrees with the time coordinate of the Rindler frame,
$\eta^{\prime}=ct^{\prime}/\xi$, the radiation remains thermal radiation with
the same temperature $T$. \ The spatial behavior shows exactly one parameter
no matter what wavelength of radiation is involved, but now has a special
wavelength \ $l_{T}=const/k_{B}T$ associated with the temperature $T$. \ The
preferred inertial frame is the frame of the box, which is also that of the
comoving reference frame. \ The preferred length is $l_{T}$ that of the
thermal radiation within the box. \ Thus, in the comoving inertial frame, at a
single time, the spectrum can indicate the presence of thermal radiation of a
single temperature $T$ and a single preferred length $l_{T}$. \ However, the
spatial randomness is \textit{unchanged} from that in the momentarily comoving
inertial frame.

\subsubsection{Time is the Crucial Aspect Between Coordinate Frames}

It is the \textit{time} behavior in the Rindler coordinate frame is so
completely different from that in an \textit{inertial} frame. \ The initial
inertial frame $S_{i}$ has all its clocks synchronized and shows one preferred
length $l_{T}=const/k_{B}T$ for thermal radiation at a single temperature $T$.
\ This single temperature $T$ is the same one indicated by the
\textit{spatial} correlations inside the accelerating box. \ However, at a
single spatial point $\left(  \xi,y,z\right)  $ in the Rindler frame, the
correlation function is above the zero-point radiation but varies with the
distance $\xi$ from the event horizon. \ 

This surprising \textit{relativistic} aspect is something which physicists
often use these days, but are frequency unaware of. \ Thus, the gps-locating
information on cell phones depends upon \textit{time} corrections which
recognize that lower clocks run slower in the earth's gravitational field.
\ However, the time corrections are so small, that people living in tall
apartment buildings are completely unaware that clocks on the ground floor run
slower than the clocks on the top floor. \ However, it is these small
relativistic corrections which are so crucial to understanding thermal radiation.

\subsection{Introduction of a Conformal-Invariant Constant into the Rindler
Frame}

At very large distances $\xi$ from the event horizon where the acceleration is
small, the coordinate time variation in the Rindler frame goes over to that of
the associate momentarily comoving inertial frame. \ However, as one goes to
ever-larger values of $\xi$ where the acceleration $a\left(  \xi\right)
=c^{2}/\xi$ \ becomes ever smaller, the temperature of the thermal radiation
decreases also, $T\left(  \xi\right)  =const/\xi$. \ \ For zero-point
radiation at $T=0$, we would simply recover the spectrum of zero-point
radiation in an inertial frame when we were very far from the event horizon.
\ However, what we would like to do is to keep the temperature constant on
moving to ever larger coordinates $\xi$ in the Rindler frame. \ Thus we will
write the temperature as $T\left(  \xi\right)  =\left[  a\left(  \xi\right)
\right]  \left[  \zeta\left(  \xi\right)  \right]  =\left[  c^{2}/\xi\right]
\left[  \zeta\left(  \xi\right)  \right]  $, where $\zeta\left(  \xi\right)  $
is some conformal-invariant parameter which \textit{changes} with distance
$\xi$ in the same manner as $\xi$. \ Thus, while changing the location $\xi$
of our box out to the region where the acceleration is very small and the
behavior is like that of an inertial coordinate frame, the temperature remains
unchanged. \ What we are really doing is using the \textit{time} correlation
function from the Rindler frame but treating the time as though it were in an
inertial frame, which it \textit{is} to a good approximation at large values
of $\xi.$ \ Thus for large values of $\xi,$ the box of radiation should behave
as though it were in an inertial frame.

The multiplication by $\zeta$ does not change the allowed normal modes for the
box, the Dirichlet boundary conditions at the walls of the box, or the
distance $\xi$ to the event horizon. \ This is still a steady-state
distribution in the Rindler frame. \ However, the new constant $\zeta$
associated with time does introduce a variable real constant $\zeta$ which is
at our disposal and should be related to the temperature. \ Thus in equation
(\ref{X}), we change the coordinate time difference $\eta^{\prime}-\eta$ over
to $\zeta\left(  \eta^{\prime}-\eta\right)  =\zeta\left[  a\left(  \xi\right)
\right]  \left(  ct^{\prime}-ct\right)  /c^{2}$ and the distance $\xi$ over to
$\zeta/\xi=\zeta\left[  a\left(  \xi\right)  /c^{2}\right]  $. \ With this new
constant $\zeta$, the equation (\ref{X}) becomes in the asymptotic large-$\xi$
region%
\begin{align}
\left\langle \phi^{\zeta}(ct,\xi,y,z)\phi^{\zeta}\left(  ct^{\prime}%
,\xi,y,z\right)  \right\rangle  &  =\frac{\hbar c}{4\pi}\left(  \frac{\zeta
}{\xi}\right)  ^{2}\frac{1}{\sinh^{2}\left[  \zeta\left(  \eta^{\prime}%
-\eta\right)  /2\right]  }\nonumber\\
&  =\frac{\hbar c}{4\pi}\left(  \frac{\zeta a}{c^{2}}\right)  ^{2}\frac
{1}{\sinh^{2}\left[  \zeta a\left(  ct^{\prime}-ct\right)  /\left(
2c^{2}\right)  \right]  }, \label{Z}%
\end{align}
where $a\equiv a\left(  \xi_{0}\right)  $ is the acceleration at the box
location $\xi_{0}$ in the Rindler frame. \ If the acceleration ceases at
$\eta_{f}$, then thereafter, the system corresponds to thermal radiation in
the $S_{f}$ inertial frame.

\subsubsection{Fourier Frequency Transform}

For the correlation function in Eq. (\ref{Z}), we would like to know the
associated frequency spectrum associated with a fixed value of $\zeta a$.
\ The Fourier transformation of the function $\omega\coth\left[  \alpha
\omega\right]  $ is\cite{G-R}%

\begin{align}
&  \int_{0}^{\infty}d\omega\left(  \omega\coth\left[  \frac{\pi c}{\zeta
a}\omega\right]  \right)  \cos\left[  \omega t\right] \nonumber\\
&  =\int_{0}^{\infty}d\omega\,\omega\cos\left[  \omega t\right]  +\int%
_{0}^{\infty}d\omega\left(  \frac{2\omega}{\exp\left[  2\pi c\omega/\left(
\zeta a\right)  \right]  -1}\right)  \cos\left[  \omega t\right] \nonumber\\
&  =-\frac{1}{t^{2}}+\left[  \frac{1}{t^{2}}-\left(  \frac{\zeta a}%
{2c}\right)  ^{2}\text{csch}^{2}\left(  \frac{\zeta a}{2c}t\right)  \right]
=-\left(  \frac{\zeta a}{2c}\right)  ^{2}\text{csch}^{2}\left(  \frac{\zeta
a}{2c}t\right)  . \label{Fourierw}%
\end{align}
Then writing csch$\left(  x\right)  =\left[  \sinh\left(  x\right)  \right]
^{-1}$, our correlation function in Eq. (\ref{Z}) becomes%

\begin{align}
\left\langle \phi^{\zeta}\left(  ct,\xi,y,z\right)  \phi^{\zeta}\left(
ct^{\prime},\xi,y,z\right)  \right\rangle  &  =\left(  \frac{\hbar}{4\pi
c^{3}}\right)  \left(  \zeta a\right)  ^{2}\frac{-1}{\sinh^{2}\left[  \zeta
a\left(  ct-ct^{\prime}\right)  /\left(  2c^{2}\right)  \right]  }\nonumber\\
&  =\left(  \frac{\hbar}{\pi c}\right)  \left(  \zeta a\right)  ^{2}\int%
_{0}^{\infty}d\omega\left(  \omega\coth\left[  \frac{\pi c}{\zeta a}%
\omega\right]  \right)  \cos\left[  \omega\left(  t-t^{\prime}\right)
\right]  .
\end{align}

\subsubsection{Connection of $\zeta a$ with Temperature}

Now we want to integrate over the finite \textit{thermal} part of the spectrum
to obtain the connection of $\zeta a$ with temperature $T$ as given in the
Stefan-Boltzmann law, $u\left(  T\right)  =a_{S}T^{4}$. \ Subtracting off the
zero-point radiation spectrum, we find%

\begin{align}
&  \int_{0}^{\infty}d\omega\frac{\omega^{2}}{\pi^{2}c^{3}}\left(  \frac{1}%
{2}\hbar\omega\coth\left[  \frac{\pi c}{\zeta a}\omega\right]  -\frac{1}%
{2}\hbar\omega\right) \nonumber\\
&  =\left(  \frac{\hbar c}{2\pi^{2}c^{4}}\right)  \left(  \frac{\zeta a}{\pi
c}\right)  ^{4}\int_{0}^{\infty}du\,u^{3}\left(  \coth\left[  u\right]
-1\right) \nonumber\\
&  =\left(  \frac{\hbar c}{2\pi^{2}c^{4}}\right)  \left(  \frac{\zeta a}{\pi
c}\right)  ^{4}\frac{1}{8}\frac{\pi^{4}}{15}=a_{S}T^{4}=\frac{\pi^{2}k_{B}%
^{4}}{15\hbar^{3}c^{3}}T^{4},
\end{align}

so that the connection with temperature is%

\begin{equation}
T=\frac{\left(  \zeta a\right)  \hbar}{2\pi ck_{B}}. \label{T}%
\end{equation}

But now we can replace the expression $\zeta a$ by the familiar temperature
$T$. \ Comparing with Eq. (\ref{Utotneed}), we have found exactly the Planck
spectrum including zero-point radiation%

\begin{align}
U^{total}\left(  \omega,T\right)   &  =U^{T}\left(  \omega,T\right)
+U^{zp}\left(  \omega\right)  =\frac{1}{2}\hbar\omega\coth\left[  \frac
{\hbar\omega}{2k_{B}T}\right] \nonumber\\
&  =\frac{\hbar\omega}{\exp\left[  \hbar\omega/\left(  k_{B}T\right)  \right]
-1}+\frac{1}{2}\hbar\omega.
\end{align}

\subsubsection{The Work-Energy Theorem for Relativistic Waves}

Since classical zero-point radiation is assumed to be universally present, it
will provide a fluctuating force on any conducting wall of a box. \ However,
it will not provide a time-average force upon the wall because the forces
inside and outside the box will cancel on time average. \ Also, there is no
external force required to accelerate zero-point radiation, because of the
zero-point radiation both inside and outside the box while the spectrum is
Lorentz invariant. \ The sum of the external forces has to provide merely the
time-rate-of-change of momentum of the walls of the box.

On the other hand, in the classical point of view adopted here, thermal
radiation is random radiation above the zero-point radiation; and thermal
energy can be in equilibrium in a box. \ In an inertial frame, the resultant
of the external forces accelerating a box of relativistic waves must provide
the change of momentum of the walls of the box plus the change of momentum
associated with the \textit{thermal} energy of the waves within the box. \ The
situations for classical zero-point energy and classical thermal energy are
very different. \ 

\subsubsection{Existence of an Asymptotic Region in the Rindler Frame}

We have assumed that we could go to an asymptotic region in the Rindler
coordinate frame where the proper time came close to the momentarily comoving
inertial frame interval. \ However, a Rindler frame is invariant under
$\sigma_{ltU^{-1}}$-dilations of the conformal group just as is Minkowski
spacetime. \ There is no preferred spacetime length or time or energy in
either a Rindler coordinate frame or a Minkowski coordinate frame. \ We must
have some parameter of length or time or energy which we can use as a
comparison in order to claim that we are in the Rindler asymptotic region
where the spacetime becomes like that of Minkowski spacetime. \ 

We notice emphatically that at \textit{any single coordinate time }$\eta$ in
the Rindler frame, the \textit{spatial} aspects are exactly those of the
momentarily comoving Minkowski inertial frame; zero-point radiation remains
zero-point radiation and thermal radiation remains thermal radiation. \ It is
rather the \textit{time} aspect which is so different between the Rindler
frame and an inertial frame. \ If there is no special length or time, the
Rindler asymptotic region \textit{can not be defined}, since the space is
invariant under the $\sigma_{ltU^{-1}}$-dilations of the conformal group.

In our classical analysis, it is \textit{the length of the box }$L$ which
provides a special comparison value. \ The ends of the box are at $\xi_{0}$
and $\xi_{0}+L$ in the direction of acceleration. \ Thus, far from the event
horizon, when the distance $\xi_{0}$ is large compared to $L$, we are in
\textit{the Rindler asymptotic region for the box}, $L/\xi_{0}<<1$. \ In the
approximation $L/\xi_{0}<<1$, corresponding to the asymptotic region, the two
spacetime dimensional Rindler normal mode in the box in Eq. (\ref{betan})
becomes approximately the Minkowski normal mode $\omega_{n}=ck_{n}$. \ Thus,
we have, for $L/\xi_{0}<<1,$ $\xi=\xi_{0}+x$%

\begin{equation}
\ln\left[  \left(  \xi_{0}+L\right)  /\xi_{0}\right]  =\ln\left[  1+\left(
L/\xi_{0}\right)  \right]  \approxeq L/\xi_{0},
\end{equation}
so that%
\begin{equation}
\sin\left[  \frac{n\pi}{\ln\left[  \left(  \xi_{0}+L\right)  /\xi_{0}\right]
}\ln\left[  \frac{\xi_{0}+x}{\xi_{0}}\right]  \right]  \approxeq\sin\left[
\frac{n\pi}{L/\xi_{0}}\left(  \frac{x}{\xi_{0}}\right)  \right]  =\sin\left[
\frac{n\pi x}{L}\right]  .
\end{equation}
\ 

\subsubsection{Determination of the Constant $\zeta$}

We see that our classical analysis depends upon having a box of finite length
$L$ in which we can define normal modes of oscillation in both Minkowski
spacetime coordinates and in Rindler spacetime coordinates. \ The presence of
thermal radiation at a temperature above zero, is required for the definition
of the constant $\zeta$. \ At zero temperature, $\left[  a(\xi)\right]
\zeta=0$ requires $\zeta=0$. \ In the asymptotic region of the Rindler
coordinate frame for our box, we still have $\left[  a(\xi)\right]  \zeta=0$.
\ We find no thermal radiation, only zero-point radiation. \ If the
temperature is above zero, then Eq. (\ref{T}) requires that $\zeta$ is not zero.

\section{Review and Comments}

In this article, we point out that acceleration using a Rindler coordinate
frame acting for a finite time provides a test for equilibrium of a random
distribution of free particles or source-free relativistic waves in a box of
finite size. \ We consider the equations of motion for particles or waves in
the box, and give derivations of the J\"{u}ttner distribution for relativistic
particles and of the Planck spectrum plus zero-point radiation for
relativistic waves.

In the present article, we emphasize that relativistic waves satisfy conformal
symmetry. \ Classical zero-point radiation is a spectrum of random classical
radiation which is conformal \textit{invariant}. \ Zero-point radiation has no
parameters which vary under a change of the standards of length, time, and
energy. \ Experimental work involving Casimir forces at small temperatures
confirms the functional dependence depending upon the zero-point radiation
spectrum, and gives a constant $\hbar/2$ required to fit the forces
quantitatively. \ Here $\hbar$ takes the same numerical value as Planck's
constant. \ Since classical zero-point radiation is the same in every inertial
frame, its correlation functions depend upon the geodesic separations between
the field points. \ On the other hand, \textit{thermal} radiation for
relativistic classical wave fields is characterized by exactly one variable
parameter, its temperature $T$, which changes under a $\sigma_{ltU^{-1}}%
$-dilation of the conformal group. \ Thus zero-point radiation and thermal
radiation are quite distinct in classical physics; the zero-point radiation is
$\sigma_{ltU^{-1}}$-dilation \textit{invariant,} and the thermal radiation is
\textit{covariant}, changing with a change in scale as, $T^{\prime}=T/\sigma$
where $\sigma$ is a positive real number. \ 

In the late 1970s, it was suggested that an observer using classical theory
and accelerating uniformly through classical zero-point radiation would find
field correlations for the vacuum which would correspond to those of thermal
radiation.\cite{B42} \ Although this effect was proposed first in quantum
field theory, it seemed to appear also in classical physics with classical
zero-point radiation.\cite{B42} \ However, the classical point of view
gradually shifted. \ It was suggested later that a box full of classical
zero-point radiation remains at zero temperature even under acceleration,
though its time correlation function may be different.\cite{Bacc} \ In this
article, we give additional support to this later point of view within
classical physics.

\ 

November 14, 2025 \ \ ConformalPlanckAcc7.tex

\end{document}